\documentclass[letterpaper, 12pt]{ewshm}

\usepackage[dvips,final]{graphicx}
\usepackage{color}
\usepackage{amsfonts}
\usepackage{amsmath}
\usepackage{amssymb}
\usepackage{times}
\usepackage{cite}
\usepackage{hhline}
\usepackage{compactbib}
\usepackage{float}

\renewcommand{\thefigure}{\arabic{figure}}
\renewcommand{\thetable}{\Roman{table}}

\usepackage[numbers]{natbib}
\usepackage{tikz}
\usetikzlibrary{calc,patterns,decorations.pathmorphing,decorations.markings}
\usetikzlibrary{arrows,calc,shapes,decorations.pathreplacing}
\usetikzlibrary{decorations.markings}
\usetikzlibrary{trees,positioning,fit,shapes}
\usetikzlibrary{backgrounds}
\usetikzlibrary{patterns, positioning, arrows}

\makeatletter
\newcommand\setveclength[3]{
  \pgfpointdiff{\pgfpointanchor{#2}{center}}{\pgfpointanchor{#3}{center}}
  \pgfmathveclen{\pgf@x}{\pgf@y}
  \edef#1{\pgfmathresult}
}
\makeatother

\usepackage{amsmath}
\usepackage{bm}
\usepackage{amsfonts}
\usepackage{xfrac}
\long\def\symbolfootnote[#1]#2{\begingroup%
\def\thefootnote{\fnsymbol{footnote}}\footnote[#1]{#2}\endgroup}

\makeatletter
\renewcommand\@biblabel[1]{#1.}
\makeatother

\begin{document}
\title{\bf On the application of topological data analysis: a Z24 Bridge case study.}	
\author{T.\ Gowdridge, N.\ Dervilis, K.\ Worden \\
        Dynamics Research Group, \\ 
        Department of Mechanical Engineering, University of Sheffield \\
        Mappin Street, Sheffield S1 3JD, UK
	   }
\date{}
\maketitle

\noindent \uppercase{\textbf{ABSTRACT}} \vspace{12pt} 

Topological methods are very rarely used in structural health monitoring (SHM), or indeed in structural dynamics generally, especially when considering the structure and topology of observed data. Topological methods can provide a way of proposing new metrics and methods of scrutinising data, that otherwise may be overlooked. In this work, a method of quantifying the shape of data, via a topic called \textit{topological data analysis} will be introduced. The main tool within topological data analysis is \textit{persistent homology}. Persistent homology is a method of quantifying the shape of data over a range of length scales. The required background and a method of computing persistent homology is briefly introduced here.

Ideas from topological data analysis are applied to a Z24 Bridge case study, to scrutinise different data partitions, classified by the conditions at which the data were collected. A metric, from topological data analysis, is used to compare between the partitions. The results presented demonstrate that the presence of damage alters the manifold shape more significantly than the effects present from temperature.

\symbolfootnote[0]{\hspace*{-7mm} \textsf{Dynamics Research Group, Department of Mechanical Engineering, University of Sheffield, Mappin St, Sheffield, S1 SJD UK}}

\vspace{24pt} 

\noindent \uppercase{\textbf{Introduction}} \vspace{12pt}

\textit{Topological data analysis} (TDA) is a recently-developed and fast-growing field that has found its way into many areas of analysis, such as Economics and Genomics \cite{economicsTDA, diabetes, genomics}. The general idea of TDA applies concepts from algebraic topology to data \cite{edelsbrunner2000topological, zomorodian2005topology}. The primary focus of TDA is to determine the shape of the manifold, from which some data are sampled. This process is achieved by identifying $2D$ holes, $3D$ cavities and higher-dimensional analogues within the data structure, over a range of length scales, by a process called \textit{persistent homology}. 

In this paper, persistent homology is applied to the Z24 bridge. Persistent homology can be used to determine the shape of some data manifold. The shape of a data manifold is determined by the conditions from which the data are sampled. Therefore, by understanding the shape of the manifold, an understanding of the system is inferred.  The Z24 bridge is a structure spanning from Koppigen to Utzenstorf, in Switzerland. A year before the dismantling of the bridge, data were collected, with a sensor network in place, to observe modal behaviour. Sensors were also used to measure the air temperature, soil temperature and humidity. Because of the extreme conditions of the Swiss weather, the air temperature was recorded as low as $-9^{\circ}C$ and as high as $36^{\circ}C$. As a result, the temperature effects are clearly visible on the calculated natural frequencies \cite{peeters2001one}. Shortly before the destruction of the bridge, there was damage introduced to the system, which is also visible in the natural frequencies after a certain point in time. The change in temperature ($\approx 30\%$) has a greater impact than introducing damage ($\approx 7\%$). For this reason, the change in the magnitude of the natural frequencies offers little insight into the presence of damage. The main problem here is to separate the damage case from the temperature effects. This paper aims to discuss how the topological structure has been altered before and after the presence of damage.

The outline of this paper is as follows: Section 2 will introduce TDA, persistent homology, and outline a brief path for conducting TDA for some sampled point data. Section 3 will introduce the Z24 problem, and apply TDA to form an argument for why manifold shape is a viable analysis technique for SHM. Section 4 will provide a few short concluding remarks of the paper.

\vspace{24pt} 
\noindent \uppercase{\textbf{TDA and Persistent Homology}} \vspace{12pt}

\begin{figure}[b]
    \centering
    
\begin{tikzpicture}[ele/.style={fill=black,circle,minimum width=3pt,inner sep=2pt}]
\node[ele] (a1) at (0,0) {};
\node at (0,-0.5) {$\Delta^0$};

\begin{scope}[shift={(2,0)}]
    \node[ele] (a1) at (0,0) {};    
    \node[ele] (a2) at (2,0) {};
    \draw[-,thick,shorten <=2pt,shorten >=2pt, red] (a1.center) -- (a2.center);
    \node at (1,-0.5) {$\Delta^1$};
    
    \node[ele] (a1) at (0,0) {};    
    \node[ele] (a2) at (2,0) {};
\end{scope}

\begin{scope}[shift={(5.5,0)}]
    \node[ele] (a1) at (2,0) {};    
    \node[ele] (a2) at (0,0) {};
    \node[ele] (a3) at (1,1.732) {};

    \draw[-,thick,shorten <=2pt,shorten >=2pt, red] (a1.center) -- (a2.center);
    \draw[-,thick,shorten <=2pt,shorten >=2pt, red] (a1.center) -- (a3.center);
    \draw[-,thick,shorten <=2pt,shorten >=2pt, red] (a2.center) -- (a3.center);

    \begin{scope}[on background layer]
        \path [fill=blue!30, draw] (a1.center) to (a2.center) to (a3.center) to (a1.center);
    \end{scope} 

    \node at (1,-0.5) {$\Delta^2$};
    \node[ele] (a1) at (2,0) {};    
    \node[ele] (a2) at (0,0) {};
    \node[ele] (a3) at (1,1.732) {};
    
\end{scope}

\begin{scope}[shift={(9,0)}]
    \node[ele] (a1) at (2,0) {};    
    \node[ele] (a2) at (0,0.132) {};
    \node[ele] (a3) at (1.32,0.66) {};
    \node[ele] (a4) at (1,2) {};
    
    \draw[-,thick,shorten <=2pt,shorten >=2pt, red] (a1.center) -- (a2.center);
    \draw[-,thick,shorten <=2pt,shorten >=2pt, red] (a1.center) -- (a4.center);
    \draw[-,thick,shorten <=2pt,shorten >=2pt, red] (a2.center) -- (a4.center);

    \draw[-,thick,shorten <=2pt,shorten >=2pt, dashed, red] (a1.center) -- (a3.center);
    \draw[-,thick,shorten <=2pt,shorten >=2pt, dashed, red] (a2.center) -- (a3.center);
    \draw[-,thick,shorten <=2pt,shorten >=2pt, dashed, red] (a4.center) -- (a3.center);
    
    \begin{scope}[on background layer]
        \path [fill=lightgray, draw] (a1.center) to (a2.center) to (a4.center) to (a1.center);
    \end{scope} 
\node at (1,-0.5) {$\Delta^3$};
    \node[ele] (a1) at (2,0) {};    
    \node[ele] (a2) at (0,0.132) {};
    \node[ele] (a3) at (1.32,0.66) {};
    \node[ele] (a4) at (1,2) {};

\end{scope}
\end{tikzpicture}
    \caption{The first four simplices.}
    \label{fig:simplices}
\end{figure}
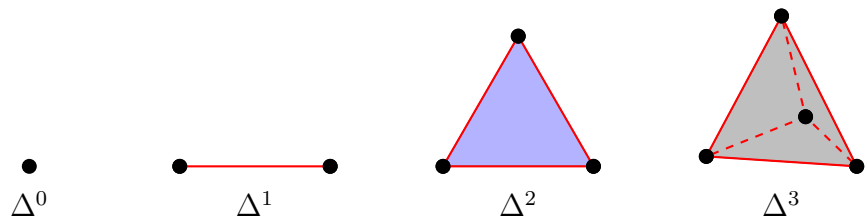

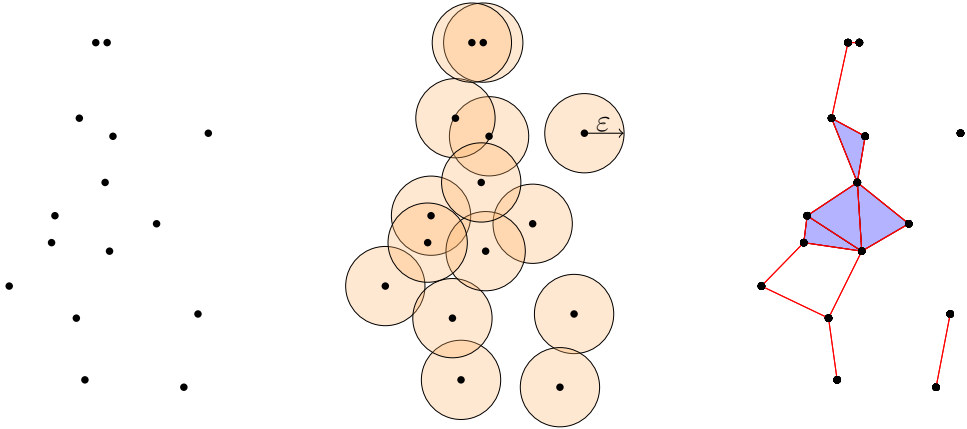
\begin{figure}[H]
    \centering
    \begin{tikzpicture}[ele/.style={fill=black,circle,minimum width=2pt,inner sep=1pt}, scale = 1]

    \node[ele] (a) at (2.8209087137275, 1.18093727264829) {};
    \node[ele] (b) at (2.95833231728538,	3.58594837328201) {};
    \node[ele] (c) at (0.311441915093690,	1.55228774215654) {};
    \node[ele] (d) at (1.69060866126193,	3.54516924716448) {};
    \node[ele] (e) at (1.31799287927655,	0.304557156170160) {};
    \node[ele] (f) at (1.61347687509477,	4.79163906483583) {};
    \node[ele] (g) at (2.27043233468693,	2.38135190156370) {};
    \node[ele] (h) at (1.20381702801476,	1.12542610389225) {};
    \node[ele] (i) at (0.919304590945695,	2.48862221210784) {};
    \node[ele] (j) at (1.64512903092519,	2.01677357394698) {};
    \node[ele] (k) at (1.24383442716840,	3.78580579686650) {};
    \node[ele] (l) at (1.58646607060451,	2.93089871716034) {};
    \node[ele] (m) at (0.874518089942222,	2.13017381291781) {};
    \node[ele] (n) at (1.46214937616206,	4.79124963787566) {};
    \node[ele] (o) at (2.63312304723148,	0.206799557528458) {};

\begin{scope}[elem/.style={circle, draw = black, line width = 0.2pt, fill = orange!60, fill opacity=0.3, minimum width = 30pt}, shift={(5,0)}]

\node[elem] (a) at (2.8209087137275, 1.18093727264829) {};
\node[elem] (b) at (2.95833231728538,   3.58594837328201) {};
\node[elem] (c) at (0.311441915093690,  1.55228774215654) {};
\node[elem] (d) at (1.69060866126193,   3.54516924716448) {};
\node[elem] (e) at (1.31799287927655,   0.304557156170160) {};
\node[elem] (f) at (1.61347687509477,   4.79163906483583) {};
\node[elem] (g) at (2.27043233468693,   2.38135190156370) {};
\node[elem] (h) at (1.20381702801476,   1.12542610389225) {};
\node[elem] (i) at (0.919304590945695,  2.48862221210784) {};
\node[elem] (j) at (1.64512903092519,   2.01677357394698) {};
\node[elem] (k) at (1.24383442716840,   3.78580579686650) {};
\node[elem] (l) at (1.58646607060451,   2.93089871716034) {};
\node[elem] (m) at (0.874518089942222,  2.13017381291781) {};
\node[elem] (n) at (1.46214937616206,   4.79124963787566) {};
\node[elem] (o) at (2.63312304723148,   0.206799557528458) {};

\node[ele] (a1) at (2.8209087137275, 1.18093727264829) {};
\node[ele] (b1) at (2.95833231728538,    3.58594837328201) {};
\node[ele] (c1) at (0.311441915093690,   1.55228774215654) {};
\node[ele] (d1) at (1.69060866126193,    3.54516924716448) {};
\node[ele] (e1) at (1.31799287927655,    0.304557156170160) {};
\node[ele] (f1) at (1.61347687509477,    4.79163906483583) {};
\node[ele] (g1) at (2.27043233468693,    2.38135190156370) {};
\node[ele] (h1) at (1.20381702801476,    1.12542610389225) {};
\node[ele] (i1) at (0.919304590945695,   2.48862221210784) {};
\node[ele] (j1) at (1.64512903092519,    2.01677357394698) {};
\node[ele] (k1) at (1.24383442716840,    3.78580579686650) {};
\node[ele] (l1) at (1.58646607060451,    2.93089871716034) {};
\node[ele] (m1) at (0.874518089942222,   2.13017381291781) {};
\node[ele] (n1) at (1.46214937616206,    4.79124963787566) {};
\node[ele] (o1) at (2.63312304723148,    0.206799557528458) {};

\node (epedge) at (3.48, 3.58594837328201) {};
\node (ep) at (3.21, 3.71) {$\varepsilon$};
\draw[->] (b.center) -- (epedge.center) ;

\end{scope}

\begin{scope}[shift={(10,0)}]
    \node[ele] (a) at (2.8209087137275, 1.18093727264829) {};
    \node[ele] (b) at (2.95833231728538,	3.58594837328201) {};
    \node[ele] (c) at (0.311441915093690,	1.55228774215654) {};
    \node[ele] (d) at (1.69060866126193,	3.54516924716448) {};
    \node[ele] (e) at (1.31799287927655,	0.304557156170160) {};
    \node[ele] (f) at (1.61347687509477,	4.79163906483583) {};
    \node[ele] (g) at (2.27043233468693,	2.38135190156370) {};
    \node[ele] (h) at (1.20381702801476,	1.12542610389225) {};
    \node[ele] (i) at (0.919304590945695,	2.48862221210784) {};
    \node[ele] (j) at (1.64512903092519,	2.01677357394698) {};
    \node[ele] (k) at (1.24383442716840,	3.78580579686650) {};
    \node[ele] (l) at (1.58646607060451,	2.93089871716034) {};
    \node[ele] (m) at (0.874518089942222,	2.13017381291781) {};
    \node[ele] (n) at (1.46214937616206,	4.79124963787566) {};
    \node[ele] (o) at (2.63312304723148,	0.206799557528458) {};
    \foreach \firstnode in {a, b, c, d, e, f, g, h, i, j, k, l, m, n, o}{%
   \foreach \secondnode in {a, b, c, d, e, f, g, h, i, j, k, l, m, n, o}{%
   \setveclength{\mydist}{\firstnode}{\secondnode}
   \pgfmathparse{\mydist < 30 ? int(1) : int(0)}
   \ifnum\pgfmathresult=1
     \draw[red] (\firstnode.center) -- (\secondnode.center);
   \fi
  }
    \node[ele] (a) at (2.8209087137275, 1.18093727264829) {};
    \node[ele] (b) at (2.95833231728538,	3.58594837328201) {};
    \node[ele] (c) at (0.311441915093690,	1.55228774215654) {};
    \node[ele] (d) at (1.69060866126193,	3.54516924716448) {};
    \node[ele] (e) at (1.31799287927655,	0.304557156170160) {};
    \node[ele] (f) at (1.61347687509477,	4.79163906483583) {};
    \node[ele] (g) at (2.27043233468693,	2.38135190156370) {};
    \node[ele] (h) at (1.20381702801476,	1.12542610389225) {};
    \node[ele] (i) at (0.919304590945695,	2.48862221210784) {};
    \node[ele] (j) at (1.64512903092519,	2.01677357394698) {};
    \node[ele] (k) at (1.24383442716840,	3.78580579686650) {};
    \node[ele] (l) at (1.58646607060451,	2.93089871716034) {};
    \node[ele] (m) at (0.874518089942222,	2.13017381291781) {};
    \node[ele] (n) at (1.46214937616206,	4.79124963787566) {};
    \node[ele] (o) at (2.63312304723148,	0.206799557528458) {};
}
\begin{pgfonlayer}{background}
\path[fill=blue!30, draw] (g.center) to (j.center) to (l.center) to (g.center); 
\path[fill=blue!30, draw] (d.center) to (k.center) to (l.center) to (d.center); 
\path[fill=blue!30, draw] (m.center) to (i.center) to (j.center) to (m.center); 
\path[fill=blue!30, draw] (l.center) to (j.center) to (i.center) to (l.center); 
\end{pgfonlayer}

\end{scope}

\end{tikzpicture}
    \caption{The process of constructing a VR complex.}
    \label{fig: Rips}
\end{figure}

The first step in the process of TDA is to construct a \textit{simplicial complex}. Simplicial complexes are used as a way of attributing quantifiable shape to the data; they can be thought of as higher-dimensional analogues of graphs. In TDA, the vertices of the simplicial complexes are the observed data points, as seen in Figure \ref{fig: Rips}. Simplicial complexes can be analysed to output the persistent homology, a key \textit{topological invariant} that can be used to describe the structure of the data. The persistent homology can then be used to compare between different data sets, by quantifying their topological structures \cite{genomics}. A simplicial complex is made up of fundamental building blocks, called \textit{simplices}. The first four simplices are shown in Figure \ref{fig:simplices}. Each vertex in the simplex is fully connected to all the other vertices and the space enclosed by the vertices is part of that simplex. For instance, $\Delta^2$ encloses a two-dimensional area, $\Delta^3$ encloses a three-dimensional volume; this can be generalised for $\Delta^k$ enclosing a $k-$dimensional space between $(k+1)$ fully connected vertices.

There are many ways to construct a simplicial complex from point data. For simplification, only one method will be discussed within this paper, the \textit{Vietoris-Rips} (VR) complex \cite{carlsson2006algebraic}. The VR complex is constructed from point data, to output a corresponding simplicial complex, which can then be analysed. For the VR complex, $VR_\varepsilon$, let $(X,\partial_X)$ be a finite metric space and $\varepsilon > 0$ be a fixed value, then \cite{chambers2010vietoris}:
\begin{enumerate}
    \item The vertices, $v \in X$, form the vertices in $VR_\varepsilon(X,\partial_X)$.
    \item A $k-$simplex is formed when $\partial_X(v_i,v_j) \leq 2\varepsilon, \ \forall i,j \leq k$ for some $\varepsilon > 0$.
\end{enumerate}
 
The process of constructing a VR complex is depicted in Figure \ref{fig: Rips}, for some randomly sampled data, and an arbitrary value of $\varepsilon$. The existence of a simplex is determined by how the balls intersect between the vertices. For a VR complex, a simplex between some vertices is formed if the Euclidean distance between the all the vertices is less than $\varepsilon$.

From the simplicial complexes, the \textit{homology groups}, $H_k(X)$, can be determined. The homology groups are invariants for the data set, $X$, where $k$ refers to the relevant dimension. Generally, the $k^{\text{th}}$ homology group encodes information about the number of $k-$dimensional holes in the data \cite{maclane2012homology, nash1988topology}. Under the rules of topology, discontinuities (voids) cannot be created or destroyed \cite{intrototopologybertmendelsen, ghrist2018homological}. Therefore, the homology can be used to categorise and compare between simplicial complexes, and by extension, data sets \cite{boissonnat2018geometric, schutz1980geometrical, ghrist2014EAT}. From the homology, the \textit{Betti numbers} are defined as the \textit{rank} of the homology groups. If the Betti numbers for two topological spaces are different, these spaces are not topologically identical. If two spaces are not topologically similar, a \textit{continuous bijective map} between the spaces does not exist.
The zeroth Betti number, $\beta_0$, is the rank of the zeroth homology group \cite{nash1988topology}, $H_0(X)$, and refers to the number of connected sets in $X$. The first Betti number, $\beta_1$, is the rank of the first homology group, $H_1(X)$, and refers to the number of non-contractible holes present in $X$. The second Betti number, $\beta_2$, refers to the number of enclosed volumes in the topological space. This analogy carries on further for higher dimensions.

\begin{figure}
    \centering
    \includegraphics[width = 0.8\textwidth]{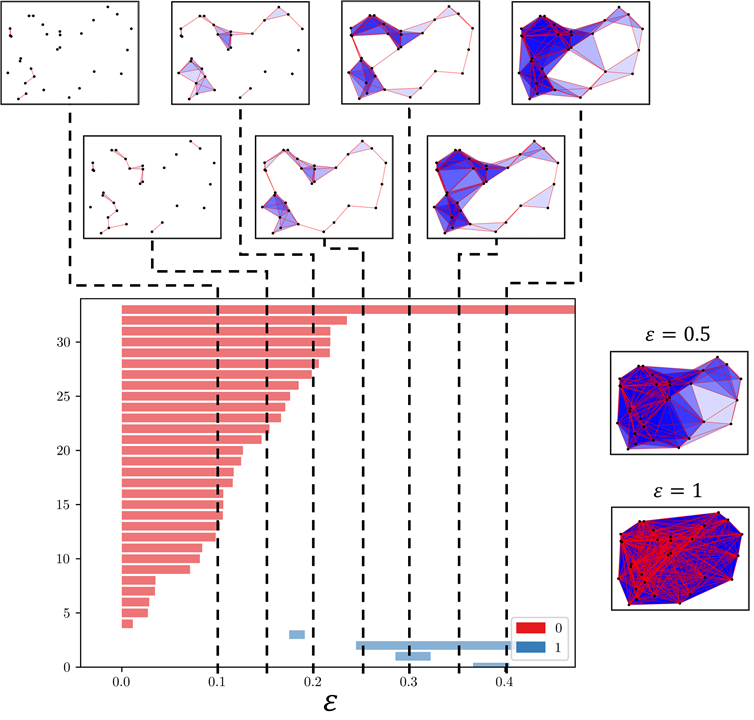}
    \caption{Persistence barcode with vertical dotted lines showing the intersections with the intervals, showing which features are present for the corresponding simplicial complex at that value of $\varepsilon$. The colour of the interval refers to the Betti number of the feature.}
    \label{fig:persistencerealisations}
\end{figure}

This now raises the question: which length scale $\varepsilon$ is representative of the topology of the data? When constructing the VR complexes, for the same data set, different values of epsilon, will result in different values for the Betti numbers. The hyper-parameter $\varepsilon$ determines the Betti numbers for that specific instance of some point cloud data. Additionally, when the feature present within the data is at a length scale less than $\varepsilon$ this feature will not be expressed, as $\varepsilon$ will span the feature. A problem arises here, as usually the feature scale is not known prior to analysis, and a manifold may have many multi-scale features. The answer to this problem, is to vary $\varepsilon$ and see how the Betti numbers evolve and \textit{persist}. Figure \ref{fig:persistencerealisations} shows some simplicial complexes in this process, for some randomly generated data. Obtaining the homology for a single value of $\varepsilon$ provides very limited information, because of potentially-varying feature length scales in the manifold. For this reason, it is vital to consider how homological features persist as $\varepsilon$ is varied. This process of varying $\varepsilon$ does not bias any disk size, as all are being considered. This process will give an initial value, $\varepsilon_{\text{min}}$, where a specific homological feature comes to life and $\varepsilon_{\text{max}}$, where the feature is no longer considered for that simplicial complex. This range of values $[\varepsilon_{\text{min}},\varepsilon_{\text{max}}]$ is called the \textit{persistence interval} for that homological feature. Each persistence interval is attributed a Betti number.

The persistence intervals obtained can be used to construct a \textit{persistence barcode}. On a barcode plot, the $x-$axis refers to the value of $\varepsilon$. As $\varepsilon$ increases, the barcode shows which features persist. The set of intervals are plotted with each interval beginning at $\varepsilon_{\text{min}}$ and ending at $\varepsilon_{\text{max}}$. The colour of the interval on the barcode refers to  the Betti number, $\beta_k$ \cite{ghrist2008barcodes}. The value of the $y-$axis can simply be thought of as an indexing of the intervals in the barcode. An example of a barcode can be seen in Figure \ref{fig:persistencerealisations}. The length of the interval represents how long the feature persists. The longer the feature persists, the higher the probability that this feature is characteristic of the manifold. Shorter intervals are generally regarded as topological noise. On the barcode, it can be seen that a feature persists to $\infty$. This is because there will always be a fully-connected simplex that persists to infinity. There will be a value of $\varepsilon_{\text{fc}}$ that results in a fully-connected simplex, where every vertex is connected to every other vertex. For values $\varepsilon>\varepsilon_{\text{fc}}$ the simplex will remain fully connected, and therefore this interval will continue to infinity.

The space of barcodes forms a \textit{metric space}; the distance between the barcodes is a measure of similarity of two barcodes. As the persistent intervals are invariant for a manifold, the data manifolds can be represented by their persistent homology. This notion of a metric space allows one to compare the similarity of manifolds. Metrics between barcodes are well established and the one used in this report is the \textit{$p-$Wasserstein distance}, $\partial_{W_p}$. 

\begin{equation}
\partial_{W_p}(B_1,B_2) = \left(\text{inf} \sum_{Z\in B_1}d_\infty(Z, \phi(Z))^p \right)^{\frac{1}{p}}    
\end{equation}

Where $B_1$ and $B_2$ are two barcodes, $p>0$ is a weighting, $\phi$ is a matching between $B_1$ and $B_2$, $Z$ is a persistence interval in $B_1$, and $d_\infty$ is the supremum metric \cite{genomics}.

\vspace{12pt} 
\noindent \uppercase{\textbf{Z24 Analysis}}  \vspace{12pt} 

\begin{figure}
    \centering
    \includegraphics[width= 0.8\textwidth]{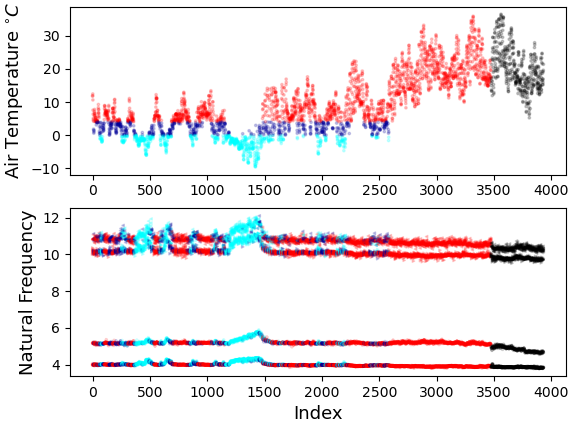}
    \caption{Above: Temperature readings plotted over the index. Below: The first four natural frequency classes, coloured by the temperature at the time of the reading.}
    \label{fig:tempnatfreq}
\end{figure}

The Z24 data set can be broken down into four categories, according to the air temperature at the time of the measurements, and whether damage was present. Figure \ref{fig:tempnatfreq} shows the temperature readings and the first four calculated natural frequencies, the corresponding colours refer to: 
\begin{enumerate}
    \item Light blue, making up the freezing data set; this is any value with a temperature reading $T < 0^{\circ}C$.
    \item Dark blue, making up the cold data set; this is any value in the temperature range $0^{\circ}C \leq T < 4^{\circ}C$.
    \item Red, making up the warm data set; this is any value with a temperature reading $4^{\circ}C \leq T$.
    \item Black, making up the damage data set; this is any reading taken after an index of $3475$, irrespective of the temperature.
\end{enumerate}

Since the data have partitioned into freezing, cold, warm, and damage, TDA can be used to compare the relative shapes of these manifolds. It is believed that the introduction of damage will change the shape of the manifold in a more substantial way, when compared to temperature changes.

By sampling over the time frame of a year, there will be slight changes in humidity, air temperature and soil temperature. These changes mean that each natural frequency will be slightly different each day. The first four natural frequencies have previously been extracted at each instance \cite{peeters2001one}. A point can be plotted in $\mathbb{R}^4$, with the $i^{\text{th}}$ axis referring to the value of $\omega_i$. Plotting these points will trace out a manifold shape that is paramaterised by the environmental and operational variations (EOVs) present over the year time frame. The data points are then assumed to lie on a manifold, representative of this specific bridge. TDA can then be used to form an understanding of what is expected for the shape of the natural frequency manifold for the Z24 data set.

The persistent homology of the data set partitions are calculated accordingly. Each data partition has a different number of points, with the warm data set being the largest partition. The warm partitions will be randomly split in half to form two additional data sets. The original warm data set will also remain in the analysis. These random subsets can be used to verify the results, as they will have a very similar topological structure, as they are all sampled from the same manifold. There will be slight differences owing to topological noise, formed from missing points in the smaller samples. These effects should be negligible over the true global structure of the manifold, when there are enough points to adequately describe the topology. For smaller partitions, the absence of many missing points will affect the topology in a significant way.

\begin{table}
    \caption{Wasserstein values over each partition.}
    \vspace{2mm}
    \centering
    \begin{tabular}{lcccccc}
    \hline
    \textit{\textbf{}}        & \textit{\textbf{Freezing}} & \textit{\textbf{Cold}} & \textit{\textbf{Warm}} & \textit{\textbf{Damage}} & \textit{\textbf{Warm1}} & \textit{\textbf{Warm2}} \\ \hline
    \textit{\textbf{Freezing}} & 0.00 & 9.39 & 22.92 & 10.62 & 12.62 & 12.50 \\
    \textit{\textbf{Cold}}     & 9.39 & 0.00 & 21.47 & 5.35 & 8.78 & 8.50 \\
    \textit{\textbf{Warm}}     & 22.92 & 21.46 & 0.00 & 23.44 & 14.26 & 14.51 \\
    \textit{\textbf{Damage}}  & 10.62 & 5.35 & 23.44 & 0.00 & 10.09 & 9.59 \\
    \textit{\textbf{Warm1}}    & 12.62 & 8.78 & 14.26 & 10.09 & 0.00 & 1.80 \\ 
    \textit{\textbf{Warm2}}    & 12.50 & 8.50 & 14.51 & 9.59 & 1.80 & 0.00\\ \hline
    \end{tabular}
    
    \label{tab:wass4d}
\end{table}

Table \ref{tab:wass4d} shows the Wasserstein distances (WDs) over the different partitions of the data set. These values are relatively uninformative, as the WDs are a factor of the number of points present in the data. This effect can be seen, as all the WDs for the full warm data set are roughly twice the size of the two random subsets; it shows that the number of points present in the point cloud is linked to the size of the Wasserstein value. It is the aim of this paper to create a metric that is independent of the number of points present in data set. As a more informative measure, the WDs can be summed along the rows. By summing along the rows, this will give an understanding of how different that data set is from all the others. The larger the value, the more topologically different the data set is. In addition, the Wasserstein sum can be normalised by the number of points present in the data set. This normalised sum now acts as a discriminating measure between manifold shapes, for manifolds with a varying number of points.

\begin{figure}[H]
    \centering
    \includegraphics[width = 0.8\textwidth]{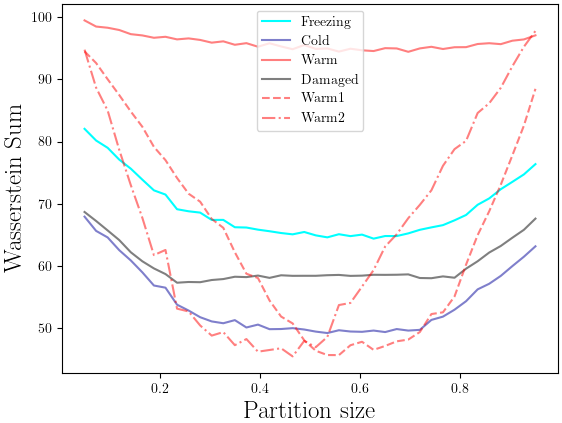}
    \caption{The size of the WDs depending on the size of the warm partition size.}
    \label{fig: varyingpartition_size}
\end{figure}

Figure \ref{fig: varyingpartition_size} shows how the sum of the WDs for any data set varies, as the partition size of the warm data set changes. It can be seen that the freezing, cold, warm, and damaged sets all change proportionally as the partitions size is varied. On the other hand, the two warm subsets vary a large amount as the partition size changes. For small partition sizes, the topology of the manifold is subsampled to an extreme, and a likeness between the manifolds cannot be established. For larger subset partition sizes, the WDs converge to the original parent data set.

\begin{figure}
    \centering
    \includegraphics[width = 0.8\textwidth]{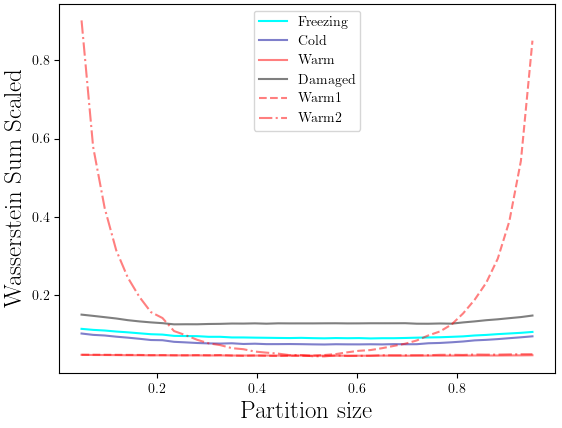}
    \caption{The size of the WD as the partition size varies, scaled by the number of points present in the data set.}
    \label{fig: varyingpartition_sizescaled}
\end{figure}
\newpage
Figure \ref{fig: varyingpartition_sizescaled} shows the sum of all the WDs for a data set, which are then scaled by the number of points present in the data set. Figure \ref{fig: varyingpartition_sizescaled} shows how dividing by the number of points almost perfectly maps the warm subsets to the full warm set, in the regions where there are enough points to adequately describe the topology.The mapping over a range of partition sizes shows that this new metric is able to distinguish between sufficiently-sampled topological spaces that contain a varying amount of data points.

As can be seen in Table \ref{tab: summedndscaledwass}, the damage case is clearly the most different in terms of manifold structure, compared to the other data sets after normalisation. The damage case is followed by freezing, cold and then the three warm data sets. This order validates the idea that the data topology is altered in a significant way when damage is introduced. This order also highlights the freezing data set as the second most different. This result is expected as a result of the material changes because of the temperature effects.

\begin{table}
\centering
\caption{Summed and scale Wasserstein data.}
\vspace{2mm}
\begin{tabular}{lccc}
    \hline
    & \textit{\textbf{Wasserstein Sum}} & \textit{\textbf{Number of Points}} & \textit{\textbf{Scaled Wasserstein Sum}} \\ \hline
    \textit{\textbf{Freezing}}   & 68.050  & 720  & 0.095\\ 
    \textit{\textbf{Cold}}    & 53.481 & 666 & 0.080 \\
    \textit{\textbf{Warm}}    & 96.584 & 2089 & 0.046 \\
    \textit{\textbf{Damage}} & 59.095 & 457 & \textbf{0.129} \\
    \textit{\textbf{Warm1}}   & 47.549 & 1044 & 0.046 \\
    \textit{\textbf{Warm2}}   & 46.903 & 1045 & 0.045 \\ \hline
\end{tabular}

\label{tab: summedndscaledwass}
\end{table}

For reference, if the random subsets are not included in the analysis, the results obtained are displayed in Tables \ref{tab: wassvalsnosubs} and \ref{tab: scalednosubs}. The damage manifold is still singled out as the most different, but this time, to a lesser extent. In this case, the damaged scenario represents a greater weight in the analysis, now that two warm subsets have been dropped. When the warm subsets were included, this gave a much larger weight to the warm condition, as three entries represented the same manifold. The warm condition can be thought of as the normal operating condition for the bridge, as this is the most frequently-observed case. Therefore, by giving extra weight to the warm condition, this helps highlight the presence of novel behaviour in the topology. Even without the extra weighting to the warm condition, the damage manifold still comes out as the most topologically dissimilar manifold. This observation indicates that the damage does change the structure of the manifold in a more significant way than the temperature changes. This result shows that topological methods can be used to distinguish the presence of damage within the system.

\begin{table}[h]
\centering
\caption{WDs with subsets not included.}
\vspace{2mm}
\begin{tabular}{lcccc}
    \hline
& \textit{\textbf{Freezing}} & \textit{\textbf{Cold}} & \textit{\textbf{Warm}} & \textit{\textbf{Damage}} \\ \hline
    \textit{\textbf{Freezing}}  & 0.00  & 9.39 & 22.92 & 10.62 \\
    \textit{\textbf{Cold}}    & 9.39  & 0.00 & 21.47 & 5.35 \\
    \textit{\textbf{Warm}}    & 22.92 & 21.46  & 0.00 & 23.44 \\
    \textit{\textbf{Damage}} & 10.62 & 5.35 & 23.44 & 0.00 \\ \hline
\end{tabular}

\label{tab: wassvalsnosubs}
\end{table}

\begin{table}[h]
\centering
\caption{Scaled summed WDs with subsets not included.}
\vspace{2mm}
\begin{tabular}{lccc}
\hline
 & \textit{\textbf{Wasserstein Sum}} & \textit{\textbf{Number of Points}} & \textit{\textbf{Scaled Wasserstein Sum}} \\ \hline
\textit{\textbf{Freezing}}   & 42.922  & 720 & 0.060 \\ 
\textit{\textbf{Cold}}    & 36.203   & 666  & 0.054 \\ 
\textit{\textbf{Warm}}    & 67.817   & 2089 & 0.032 \\ 
\textit{\textbf{Damage}} & 39.411 & 457 & \textbf{0.086}  \\ \hline
\end{tabular}

\label{tab: scalednosubs}
\end{table}

\vspace{24pt}
\noindent \uppercase{\textbf{Conclusion}} \vspace{12pt}

With respect to the Z24 data set, topological methods have been able to single out the damage data partition as the most topologically dissimilar. However, further analysis on topological methods for damage detection would need to be explored to understand the true limits and possibilities of TDA in SHM. An insight into the data structure provides powerful insight into the operating conditions of a machine or structure.

Future work on TDA, will look at different topological methods; aside from using the Wasserstein distance as a metric, other case studies will also be considered. A further journal paper, which extends on the ideas presented here is currently under submission.

\vspace{24pt}
\noindent \uppercase{\textbf{Acknowledgements}} \vspace{12pt}

The authors gratefully acknowledge the support of the UK Engineering and Physical Sciences Research Council (EPSRC) through Grant reference EP/R003645/1.

\small 
\newpage
\bibliographystyle{iwshm}
\bibliography{ref}

\end{document}